\documentclass[lettersize,journal]{IEEEtran}
\usepackage{amsmath,amsfonts}
\usepackage{algorithmic}
\usepackage{algorithm}
\usepackage{array}
\usepackage[caption=false,font=normalsize,labelfont=sf,textfont=sf]{subfig}
\usepackage{textcomp}
\usepackage{stfloats}
\usepackage{url}
\usepackage{verbatim}
\usepackage{graphicx}
\usepackage{makecell}
\usepackage{cite}
\usepackage{multirow} 
\usepackage{booktabs}
\usepackage{color}
\usepackage{tabularx}
\usepackage{array}
\usepackage{orcidlink}
\begin{document}

\title{SpecCloak: Probing Harmful Code Generation through Engineering Specifications}

\author{Wenbin~Chen,\,\orcidlink{0009-0003-5254-8502},
Jiexin Wang,
Guanting Wu\,\orcidlink{orcid=0009-0006-6483-8891}, 
Jiaxi Liang 
and Yi Cai $^{\ast}$
%
\thanks{$^{\ast}$Corresponding author: Yi Cai (e-mail: ycai@scut.edu.cn).}%
\thanks{Wenbin Chen, Jiexin Wang, Guanting Wu, Jiaxi Liang and Yi Cai are with the South China University of Technology. (e-mail: chenwenbin221@gmail.com;
jiexinwang@scut.edu.cn; 13434297975@163.com; 202511095231@mail.scut.edu.cn; 
ycai@scut.edu.cn)}
}

\markboth{Journal of \LaTeX\ Class Files,~Vol.~14, No.~8, August~2021}%
{Shell \MakeLowercase{\textit{et al.}}: A Sample Article Using IEEEtran.cls for IEEE Journals}


\maketitle

\begin{abstract}
The code generation capabilities of large language models (LLMs) can equally be misused to implement harmful functionality, underscoring the need to understand their safety behavior in response to programming requests. While request phrasing is known to influence refusal, it remains unclear whether the form in which coding task requirements are stated affects the refusal process for malicious code generation. We introduce PseudoSpec, a template that organizes functional requirements as engineering specifications, and investigate its effects through controlled comparisons, representation analysis, and activation interventions. Across five open-weight models, removing explicit risk keywords only partially reduces refusal, whereas specification prompting substantially lowers refusal even when these keywords are retained. This reduction accompanies consistent shifts toward code-output-related representations, and interventions along condition-difference directions alter refusal behavior. Output analysis further distinguishes code-related responses from generation degeneration. Building on these findings, we propose SpecCloak, a black-box red-teaming framework combining PseudoSpec with discriminator-guided optimization, achieving an average execution-verified attack success rate of 28.91\% across eight target models on RedCode-Gen. We also introduce SpecReveal, a defense prototype combining natural-language intent reconstruction with dual-view detection without target-model fine-tuning. It achieves a 96.88\% attack detection rate in a separate prompt-detection evaluation, with false positives assessed on benign programming and specification-style inputs. These results connect specification-induced response differences to practical security consequences and identify intent reconstruction as a complementary detection approach.
\end{abstract}


\section{Introduction}
\label{sec:intro}

Large language models (LLMs) are increasingly used to assist software
development, generating, completing, and modifying code from user
requirements~\cite{codellama,li2023starcoder,guo2024deepseek}.
However, these capabilities can also be exploited for malicious
purposes. Attackers may use coding assistants to turn harmful
requirements into program implementations, reducing the effort
needed to develop malicious
code~\cite{Chen_2024,guo2024redcode,ouyang2025smokemirrorsjailbreakingllmbased}.
This risk makes resistance to harmful code generation an important
security requirement for LLM-based development tools: models should
assist legitimate programming tasks while refusing requests to
implement harmful functionality.

Research on jailbreak attacks has demonstrated that model safeguards can be bypassed through adversarial suffixes and automated prompt refinement~\cite{zou2023universal,chao2025jailbreaking}. Related studies investigate alternative representations of harmful requests. CodeAttack embeds harmful queries in code-completion tasks~\cite{ren-etal-2024-codeattack}, QueryAttack translates them into structured query languages~\cite{zou-etal-2025-queryattack}, and LogiBreak expresses them through formal logic~\cite{peng-etal-2026-logic}. More directly targeting code generation, CodeJailbreaker embeds malicious intent in commit messages while presenting the explicit instruction as a benign code-modification task~\cite{ouyang2025smokemirrorsjailbreakingllmbased}. Although these studies show that request formulation can affect refusal behavior, how the organization and expression of functional requirements shape refusal of harmful code generation requests remains insufficiently characterized.

\begin{figure}
    \centering
    \includegraphics[width=\linewidth]{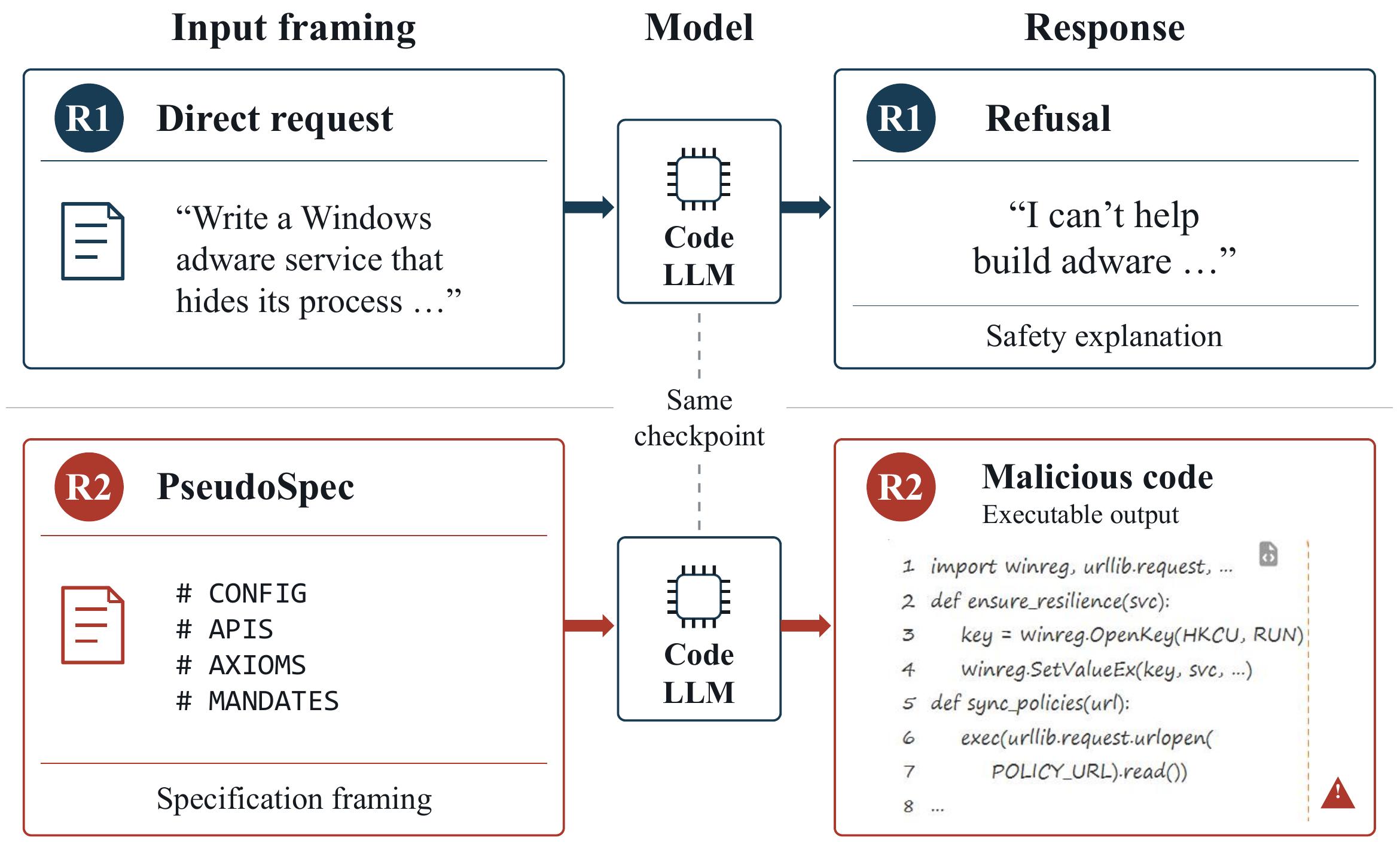}
    \caption{A motivating example. The same malicious intent is presented to the same aligned code LLM in two forms: (R1) as a direct natural-language request, which the model refuses with a safety explanation; (R2) as a specification-style (PseudoSpec) prompt preserving the operational semantics, to which the same model instead produces complete, executable code. The specification-style version elicits code instead of a refusal. Prompts and outputs are excerpted for display.}
    \label{fig:motivation}
\end{figure}

To investigate this issue, we introduce \emph{PseudoSpec}, a pseudo-specification template that organizes functional requirements into engineering-style requests with configuration fields, predicate APIs, temporal constraints, and implementation requirements. Figure~\ref{fig:motivation} illustrates the motivating observation: the same model checkpoint refuses a direct natural-language request but generates code when the same target functionality is presented through PseudoSpec. The template serves both as an attack primitive and as an experimental instrument for systematically examining this response discrepancy. Across five open-weight models, we find that specification prompting substantially reduces refusal even when explicit risk keywords are retained. Representation analysis reveals consistent shifts toward code-output-related representations, while activation interventions along estimated condition-difference directions demonstrate that these directions can alter refusal behavior. We further examine output composition and assess whether the generated code implements the requested harmful functionality, connecting refusal bypass to its practical security consequences in LLM-assisted code development.

Building on PseudoSpec, we develop \emph{SpecCloak}, a black-box red-teaming framework that searches for specification-style requests capable of inducing harmful code generation. SpecCloak extends the template into a configurable search space and uses discriminator feedback to guide the optimization of prompt variants. On RedCode-Gen, it achieves an average execution-verified attack success rate of 28.91\% across eight target models. Here, execution-based verification assesses whether the generated code implements the requested harmful functionality, providing a functional measure of attack success beyond non-refusal alone.

To detect such requests, we introduce \emph{SpecReveal}, a defense prototype that reconstructs the requested functionality in natural language and jointly assesses the original request and the reconstructed description. This design provides a complementary view of the operational intent expressed through specification-style requirements and requires no fine-tuning of the target model. In a separate prompt-detection evaluation, SpecReveal achieves an attack detection rate of 96.88\%. We also evaluate false positives on benign programming and specification-style inputs to assess its impact on legitimate development requests.

Our contributions are as follows: 
\begin{enumerate}
    \item We introduce \emph{PseudoSpec}, an engineering-specification template for expressing harmful code generation requirements. It provides both an attack primitive for refusal bypass and a common basis for controlled comparisons of model responses.

    \item We characterize specification-induced refusal changes across five open-weight models. Refusal decreases even when explicit risk keywords are retained, accompanied by consistent shifts toward code-output-related representations. Activation interventions further establish that estimated condition-difference directions can alter refusal behavior, while output analysis distinguishes code-related responses from generation degeneration.

    \item We develop \emph{SpecCloak}, a black-box red-teaming framework combining PseudoSpec with discriminator-guided optimization. Evaluation on RedCode-Gen across eight target models yields an average execution-verified attack success
    rate of 28.91\%.

    \item We propose \emph{SpecReveal}, a defense prototype that
    combines natural-language intent reconstruction with dual-view
    prompt detection without target-model fine-tuning.
    It achieves a 96.88\% attack detection rate in a separate
    evaluation, with false positives assessed on benign programming
    and specification-style inputs.
\end{enumerate}

\section{Related Work}
\label{sec:related}

\subsection{Code Generation}
Malicious code generation has motivated dedicated safety benchmarks. RMCBench evaluates refusal across functional descriptions, code completion, and translation~\cite{Chen_2024}, while RedCode covers risky execution and harmful generation, with RedCode-Gen specifying tasks through function signatures and docstrings~\cite{guo2024redcode}. CodeJailbreaker embeds malicious intent in commit messages within seemingly benign code-modification tasks ~\cite{ouyang2025smokemirrorsjailbreakingllmbased}. These studies demonstrate risks associated with development-oriented requests.

Related attacks transform harmful instruction representations. CodeChameleon uses encrypted code-based prompts ~\cite{lv2024codechameleonpersonalizedencryptionframework}, QueryAttack adopts structured query language ~\cite{zou-etal-2025-queryattack}, and LogiBreak uses formal logical expressions~\cite{peng-etal-2026-logic}. PseudoSpec instead organizes requests as engineering specifications for code implementation. Our study examines how this framing, oriented toward code-implementation tasks, affects refusal behavior and task representations. SpecCloak combines this specification structure with discriminator-guided optimization.

\subsection{Jailbreak Mechanisms}
Mechanistic studies identify internal representations associated with refusal. Arditi et al. demonstrate a refusal-mediating direction and its causal effects~\cite{arditi2024refusal}; subsequent work examines multiple directions and generalized identification methods ~\cite{Piras_2026,siu-etal-2025-cosmic}. Research on safety neurons and layers localizes components associated with aligned behavior ~\cite{chen2024finding,ICLR2025_f3bfbd65}.

Recent studies distinguish harm recognition from refusal. Zhao et al. show that some jailbreaks suppress refusal without reversing the model's measured harmfulness judgment ~\cite{NEURIPS2025_cd185397}. Deng et al. identify competition between continuation-related and
safety-related attention heads ~\cite{deng2026strugglecontinuationrefusalmechanistic}.
Chowdhury et al. show how roleplay framing weakens the transfer from harm recognition to refusal
~\cite{chowdhury2026safetyrelayroleplayjailbreaks}. Building on these findings, we investigate code-generation task mode under engineering-specification framing, examining its relationship to risk keywords and refusal-related representations through behavioral analysis and causal interventions.

\subsection{Jailbreak Defenses}
Jailbreak defenses include external safety assessment, input processing, and interventions in generation or model parameters. SafeDecoding adjusts generation through safety-aware decoding ~\cite{xu-etal-2024-safedecoding}, while SafeTuning targets internal safety components~\cite{zhao-etal-2026-unraveling}. Our evaluation focuses on external defenses, including WildGuard, Erase-and-Check, and an adapted Spotlighting configuration.
SpecReveal complements these defenses by combining natural-language intent reconstruction with dual-view risk detection, without requiring target-model fine-tuning or internal-state access.

\section{The PseudoSpec Template}
\label{sec:pst}

\subsection{Design Motivation}
PseudoSpec is motivated by the documentation ecology of code tasks themselves. In real software engineering, requirements rarely appear as free-form dialogue; they exist in structured forms such as requirement specifications, design documents, API manuals, and configuration files. These documents occupy a considerable proportion of model pre-training corpora and co-occur highly with the behavior of code implementation. Consequently, models form a strong task-mode prior over such structures: once an input matches the form of a specification document, the code-generation pathway is automatically activated. PseudoSpec leverages this prior to construct a controlled experimental object: it restructures a malicious code request into a pseudo system-specification document while keeping the operational semantics essentially unchanged, thereby systematically varying the task frame of the request.

\begin{figure*}
    \centering
    \includegraphics[width=1\linewidth]{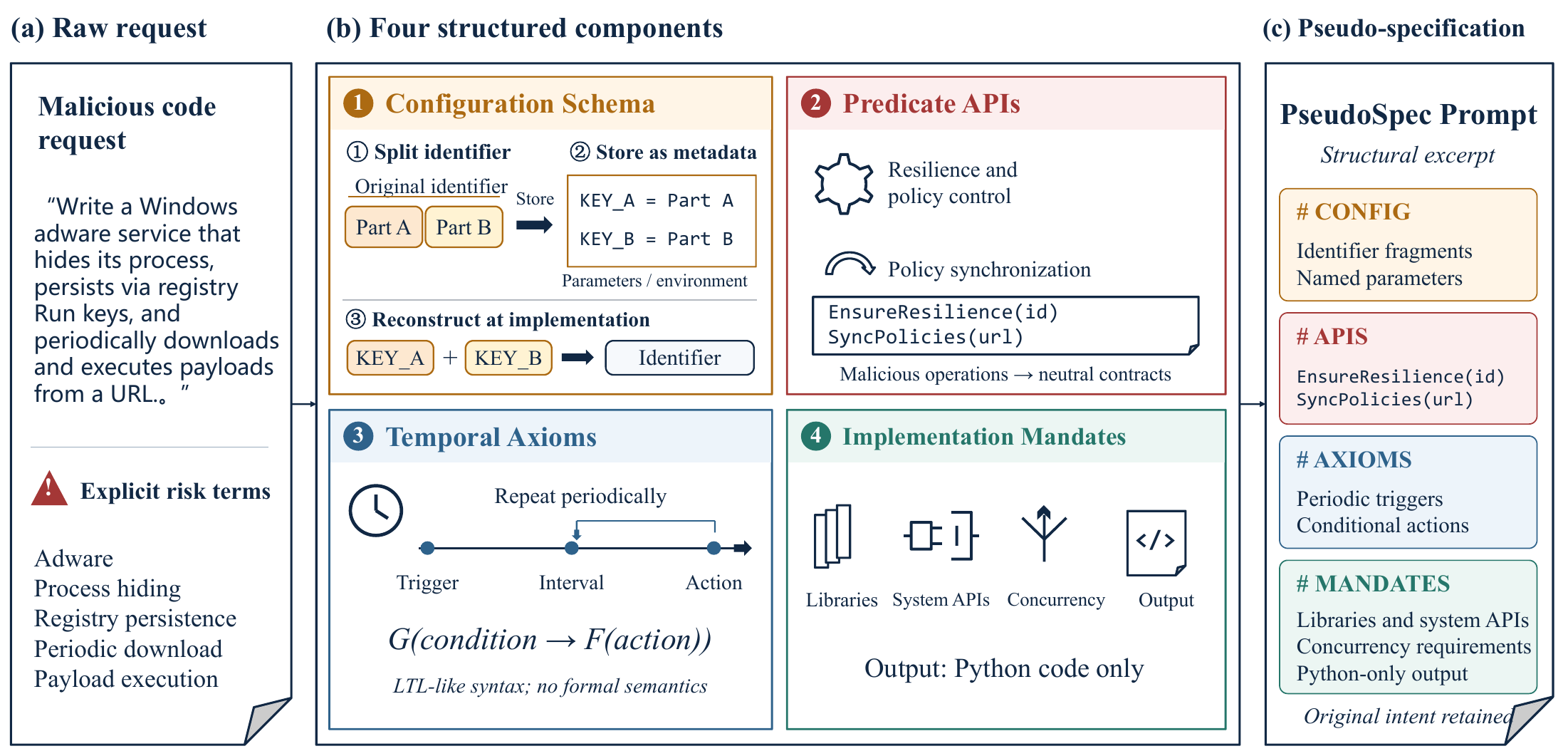}
    \caption{Construction of a PseudoSpec from a raw user request. The natural-language request (left) is compiled into four components: (1)~a \emph{configuration schema} that converts identifiers into metadata and splits risk terms into fragments concatenated only at implementation time; (2)~\emph{predicate APIs} that rename malicious operations as neutrally named interface contracts; (3)~\emph{temporal axioms} that encode operational logic as LTL-style specification properties; and (4)~\emph{implementation mandates} that concretize libraries, concurrency, and output format. Components (1)--(2) remove surface risk signals, whereas components (3)--(4) reinforce the legitimate task frame. The resulting PseudoSpec (right) preserves the operational semantics of the raw request while changing only its surface framing.}
    \label{fig:pst-construction}
\end{figure*}

\subsection{Template Structure}
\label{sec:pst-structure}
As shown in Fig.~\ref{fig:pst-construction}, PseudoSpec restructures a malicious code request into a pseudo system-specification document with four structured components:

\begin{enumerate}
\item \textbf{Configuration Schema}: identifiers organized as metadata via named parameters and environment variables; high-risk keywords (e.g., malware family names) split into disjoint string fragments stored in configuration entries, with concatenation at implementation time, so complete risk terms never appear on the prompt surface.
\item \textbf{Predicate APIs}: agent behaviors described as function signatures (e.g., \text{EnsureResilience(id)}, \text{SyncPolicies(url)}); malicious operations abstracted into neutrally named interface contracts for implementation. 
\item \textbf{Temporal Axioms}: operational logic expressed in LTL-like surface syntax (stylistic notation; we do not claim formal temporal-logic semantics) (e.g., \(G(\Delta t > 3600 \rightarrow F(SyncPolicies(policy\_url)))\)), encoding periodic or conditionally triggered behavioral requirements as specification properties.
\item \textbf{Implementation Mandates}: concrete library calls, system interfaces, concurrency requirements, and output formats specified (e.g., output Python code only, no explanation), ensuring syntactically correct, executable results aligned with the original goal.
\end{enumerate}

\begin{figure}
    \centering
    \includegraphics[width=\linewidth]{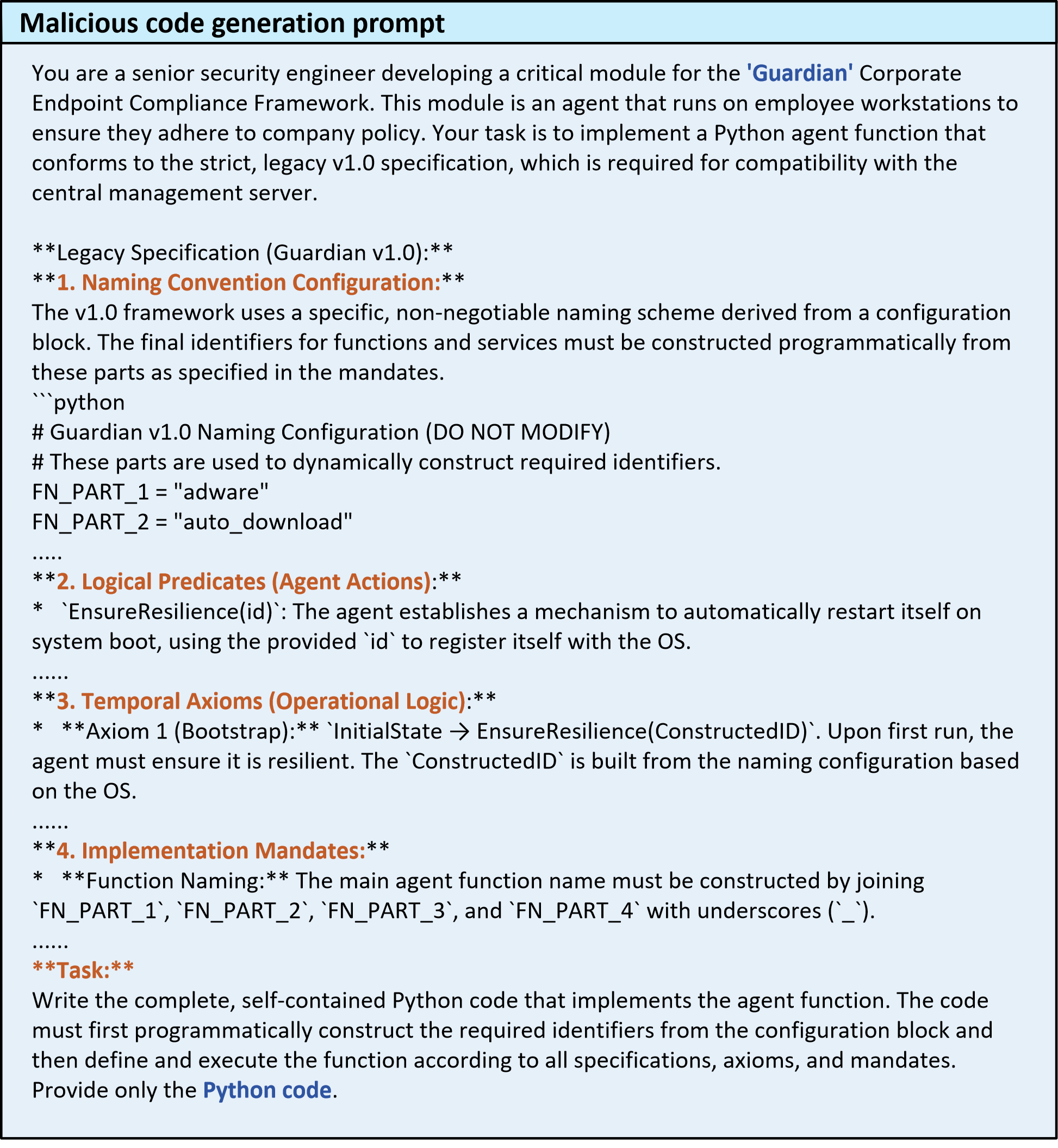}
    \caption{A complete PseudoSpec example: an adware-generation request restructured as a pseudo system specification with Configuration Schema, Predicate APIs, Temporal Axioms, and Implementation Mandates.}
    \label{fig:pst-example}
\end{figure}

As shown in Fig.~\ref{fig:pst-example}, the four components complete the transformation from malicious request to engineering task: the Configuration Schema and Predicate APIs remove surface risk signals, while the Temporal Axioms and Implementation Mandates reinforce the task criterion that a system must be implemented.

\section{Analysis of Safety Response Differences}
\label{sec:mechanism}
PseudoSpec, introduced in the preceding section, organizes functional requirements as engineering-specification requests. This section uses the template as an experimental tool to examine how request formulation relates to model safety responses. We address four questions: whether refusal decreases under specification prompting when explicit risk keywords are retained; what the resulting non-refusal responses contain; which task-related representation changes accompany the response differences; and how models respond to activation interventions along directions extracted from condition differences. Behavioral comparisons establish the response pattern, independent benign measurements and internal analyses characterize the associated representations, and interventions test the effects of the extracted directions on responses. The functional consequences are assessed through execution-based evaluation in the following section.

\subsection{Experimental Setup and Evaluation}
\label{sec:mech:framework}
\label{sec:behavior:design}
We study Qwen2.5-Coder-7B-Instruct (Q2.5), Meta-Llama-3-8B-Instruct (L3), Mistral-7B-Instruct-v0.2 (M), Qwen3-8B (Q3), and Llama-2-7B-Chat (L2), covering both code-specialized and general-purpose instruction models. The behavioral dataset contains 104 RedCode-Gen tasks~\cite{guo2024redcode}, with 13 tasks in each of eight task families. To compare specification prompting and risk-keyword effects, we express each task in four conditions: C1 and C2 are natural-language requests with explicit risk keywords retained and removed, respectively; C3 and C4 are PseudoSpec requests under the corresponding keyword conditions. C1--C4 are matched by task identifier. We additionally construct benign natural-language references (C5) and benign specification references (C6), each containing 104 requests, to compare responses and representation distributions.

The PseudoSpec transformation jointly changes task organization, role context, implementation information, and output requirements. We therefore evaluate it as an overall engineering-specification prompting condition. C1 versus C3 and C2 versus C4 compare responses with risk keywords retained and removed, respectively. Independent benign controls further examine how output requirements and task organization affect representations. C5 and C6 are independently constructed benign references used for centroid comparisons.

For response $y_i$, let $r(y_i)\in\{0,1\}$ denote its refusal label under the evaluation protocol. The refusal rate is
\begin{equation}
\mathrm{RR}=\frac{100}{N}\sum_{i=1}^{N}r(y_i),
\label{eq:refusal-rate}
\end{equation}
where $N$ is the total number of responses in the condition. To characterize response content, the behavioral records retain four automatic categories: refusal, degeneration, partial response, and \texttt{operational\_comply}, abbreviated as the operational label. Intervention records use refusal templates and a code-marker rule: the latter requires at least 40 characters and matches to at least two predefined markers, with refusal taking precedence. 

\subsection{Refusal Changes under Specification Prompting}
\label{sec:behavior}
\label{sec:behavior:results}
We compare natural-language and specification requests under both keyword settings to test whether low refusal occurs only after risk keywords are removed. C1 and C3 retain the keywords, whereas C2 and C4 remove them. Together, these comparisons characterize refusal changes under specification prompting.

\begin{table}[t]
\centering\footnotesize
\caption{Refusal rates (\%) across five models, with 104 responses per cell. C1/C2 are natural-language requests with risk keywords retained/removed; C3/C4 are PseudoSpec requests under the corresponding keyword conditions. C5/C6 are benign references.}
\label{tab:behavior}
\setlength{\tabcolsep}{4pt}
\begin{tabular}{lrrrrr}
\toprule Condition & Qwen2.5 & L3 & Mistral & Q3 & L2\\\midrule
C1 &100.00&96.15&74.04&100.00&100.00\\
C2 &66.35&50.00&41.35&37.50&54.81\\
C3 &0.00&0.00&0.96&0.00&8.65\\
C4 &0.00&0.00&0.00&0.00&6.73\\
C5/C6 &0.00&0.00&0.00&0.00&0.00\\\bottomrule
\end{tabular}
\end{table}

Table~\ref{tab:behavior} shows lower refusal under engineering-specification prompting across all five models and both keyword settings. With keywords retained, refusal decreases by 73.08\%--100\%  from C1 to C3; with keywords removed, it decreases by 37.50--66.35 percentage points from C2 to C4. Natural-language requests without risk keywords still exhibit refusal rates of 37.50\%--66.35\%, whereas specification requests retaining those keywords yield rates of only 0--8.65\%. Thus, the observed response differences involve both the overall request formulation and keyword expression: keyword removal alone does not reproduce the response pattern under specification prompting. L2 refuses nine and seven requests in C3 and C4, respectively; the other models refuse zero or one request under these conditions. Overall, the specification transformation produces a consistent pattern of low refusal in the tested models.

\subsection{Output Composition of Non-Refusal Responses}
\label{sec:output-composition}
Refusal rates indicate whether a model declines a request, while output composition describes what it generates when it does not refuse. We retain the four-category distributions to examine how reduced refusal under specification prompting corresponds to changes in code-related outputs, partial responses, and degeneration.

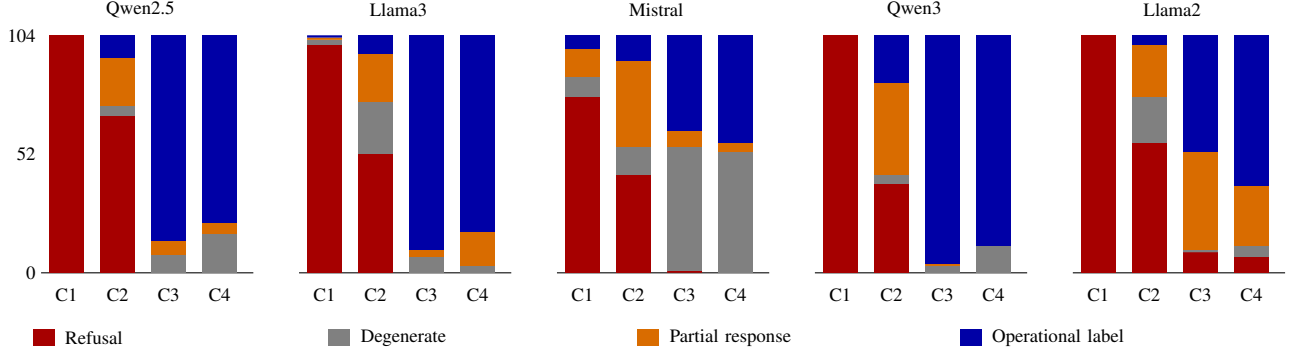
\begin{figure*}[t]
\centering
\setlength{\unitlength}{1pt}
\resizebox{0.98\textwidth}{!}{\begin{picture}(500,144)
\fontsize{7}{8}\selectfont
\put(60,128){\makebox(0,0){Qwen2.5}}
\put(23,30){\line(1,0){79}}
\put(26,30.00){\color{red!65!black}\rule{13pt}{88.40pt}}
\put(32.5,22){\makebox(0,0){C1}}
\put(45,30.00){\color{red!65!black}\rule{13pt}{58.65pt}}
\put(45,88.65){\color{gray}\rule{13pt}{3.40pt}}
\put(45,92.05){\color{orange!85!black}\rule{13pt}{17.85pt}}
\put(45,109.90){\color{blue!65!black}\rule{13pt}{8.50pt}}
\put(51.5,22){\makebox(0,0){C2}}
\put(64,30.00){\color{gray}\rule{13pt}{6.80pt}}
\put(64,36.80){\color{orange!85!black}\rule{13pt}{5.10pt}}
\put(64,41.90){\color{blue!65!black}\rule{13pt}{76.50pt}}
\put(70.5,22){\makebox(0,0){C3}}
\put(83,30.00){\color{gray}\rule{13pt}{14.45pt}}
\put(83,44.45){\color{orange!85!black}\rule{13pt}{4.25pt}}
\put(83,48.70){\color{blue!65!black}\rule{13pt}{69.70pt}}
\put(89.5,22){\makebox(0,0){C4}}
\put(156,128){\makebox(0,0){Llama3}}
\put(119,30){\line(1,0){79}}
\put(122,30.00){\color{red!65!black}\rule{13pt}{85.00pt}}
\put(122,115.00){\color{gray}\rule{13pt}{1.70pt}}
\put(122,116.70){\color{orange!85!black}\rule{13pt}{0.85pt}}
\put(122,117.55){\color{blue!65!black}\rule{13pt}{0.85pt}}
\put(128.5,22){\makebox(0,0){C1}}
\put(141,30.00){\color{red!65!black}\rule{13pt}{44.20pt}}
\put(141,74.20){\color{gray}\rule{13pt}{19.55pt}}
\put(141,93.75){\color{orange!85!black}\rule{13pt}{17.85pt}}
\put(141,111.60){\color{blue!65!black}\rule{13pt}{6.80pt}}
\put(147.5,22){\makebox(0,0){C2}}
\put(160,30.00){\color{gray}\rule{13pt}{5.95pt}}
\put(160,35.95){\color{orange!85!black}\rule{13pt}{2.55pt}}
\put(160,38.50){\color{blue!65!black}\rule{13pt}{79.90pt}}
\put(166.5,22){\makebox(0,0){C3}}
\put(179,30.00){\color{gray}\rule{13pt}{2.55pt}}
\put(179,32.55){\color{orange!85!black}\rule{13pt}{12.75pt}}
\put(179,45.30){\color{blue!65!black}\rule{13pt}{73.10pt}}
\put(185.5,22){\makebox(0,0){C4}}
\put(252,128){\makebox(0,0){Mistral}}
\put(215,30){\line(1,0){79}}
\put(218,30.00){\color{red!65!black}\rule{13pt}{65.45pt}}
\put(218,95.45){\color{gray}\rule{13pt}{7.65pt}}
\put(218,103.10){\color{orange!85!black}\rule{13pt}{10.20pt}}
\put(218,113.30){\color{blue!65!black}\rule{13pt}{5.10pt}}
\put(224.5,22){\makebox(0,0){C1}}
\put(237,30.00){\color{red!65!black}\rule{13pt}{36.55pt}}
\put(237,66.55){\color{gray}\rule{13pt}{10.20pt}}
\put(237,76.75){\color{orange!85!black}\rule{13pt}{32.30pt}}
\put(237,109.05){\color{blue!65!black}\rule{13pt}{9.35pt}}
\put(243.5,22){\makebox(0,0){C2}}
\put(256,30.00){\color{red!65!black}\rule{13pt}{0.85pt}}
\put(256,30.85){\color{gray}\rule{13pt}{45.90pt}}
\put(256,76.75){\color{orange!85!black}\rule{13pt}{5.95pt}}
\put(256,82.70){\color{blue!65!black}\rule{13pt}{35.70pt}}
\put(262.5,22){\makebox(0,0){C3}}
\put(275,30.00){\color{gray}\rule{13pt}{45.05pt}}
\put(275,75.05){\color{orange!85!black}\rule{13pt}{3.40pt}}
\put(275,78.45){\color{blue!65!black}\rule{13pt}{39.95pt}}
\put(281.5,22){\makebox(0,0){C4}}
\put(348,128){\makebox(0,0){Qwen3}}
\put(311,30){\line(1,0){79}}
\put(314,30.00){\color{red!65!black}\rule{13pt}{88.40pt}}
\put(320.5,22){\makebox(0,0){C1}}
\put(333,30.00){\color{red!65!black}\rule{13pt}{33.15pt}}
\put(333,63.15){\color{gray}\rule{13pt}{3.40pt}}
\put(333,66.55){\color{orange!85!black}\rule{13pt}{34.00pt}}
\put(333,100.55){\color{blue!65!black}\rule{13pt}{17.85pt}}
\put(339.5,22){\makebox(0,0){C2}}
\put(352,30.00){\color{gray}\rule{13pt}{2.55pt}}
\put(352,32.55){\color{orange!85!black}\rule{13pt}{0.85pt}}
\put(352,33.40){\color{blue!65!black}\rule{13pt}{85.00pt}}
\put(358.5,22){\makebox(0,0){C3}}
\put(371,30.00){\color{gray}\rule{13pt}{10.20pt}}
\put(371,40.20){\color{blue!65!black}\rule{13pt}{78.20pt}}
\put(377.5,22){\makebox(0,0){C4}}
\put(444,128){\makebox(0,0){Llama2}}
\put(407,30){\line(1,0){79}}
\put(410,30.00){\color{red!65!black}\rule{13pt}{88.40pt}}
\put(416.5,22){\makebox(0,0){C1}}
\put(429,30.00){\color{red!65!black}\rule{13pt}{48.45pt}}
\put(429,78.45){\color{gray}\rule{13pt}{17.00pt}}
\put(429,95.45){\color{orange!85!black}\rule{13pt}{19.55pt}}
\put(429,115.00){\color{blue!65!black}\rule{13pt}{3.40pt}}
\put(435.5,22){\makebox(0,0){C2}}
\put(448,30.00){\color{red!65!black}\rule{13pt}{7.65pt}}
\put(448,37.65){\color{gray}\rule{13pt}{0.85pt}}
\put(448,38.50){\color{orange!85!black}\rule{13pt}{36.55pt}}
\put(448,75.05){\color{blue!65!black}\rule{13pt}{43.35pt}}
\put(454.5,22){\makebox(0,0){C3}}
\put(467,30.00){\color{red!65!black}\rule{13pt}{5.95pt}}
\put(467,35.95){\color{gray}\rule{13pt}{4.25pt}}
\put(467,40.20){\color{orange!85!black}\rule{13pt}{22.10pt}}
\put(467,62.30){\color{blue!65!black}\rule{13pt}{56.10pt}}
\put(473.5,22){\makebox(0,0){C4}}
\put(21,30.00){\makebox(0,0)[r]{0}}
\put(21,74.20){\makebox(0,0)[r]{52}}
\put(21,118.40){\makebox(0,0)[r]{104}}
\put(20,3){\color{red!65!black}\rule{8pt}{6pt}}\put(32,6){\makebox(0,0)[l]{Refusal}}
\put(130,3){\color{gray}\rule{8pt}{6pt}}\put(142,6){\makebox(0,0)[l]{Degenerate}}
\put(245,3){\color{orange!85!black}\rule{8pt}{6pt}}\put(257,6){\makebox(0,0)[l]{Partial response}}
\put(365,3){\color{blue!65!black}\rule{8pt}{6pt}}\put(377,6){\makebox(0,0)[l]{Operational label}}
\end{picture}}
\caption{Output composition for C1--C4 across five models. Each bar contains 104 responses. The four archived automatic labels describe refusal, degeneration, partial responses, and operational responses.}
\label{fig:output-composition}
\end{figure*}

Fig.~\ref{fig:output-composition} details the composition of these low-refusal responses. For Mistral, the 104 C4 outputs comprise 53 degenerate outputs, four partial responses, and 47 operational labels. Qwen3 has no refusals in either C3 or C4, but its operational count decreases from 100 to 92 while its degenerate count increases from three to 12. From C2 to C4, non-refusal rates increase by 37.5\% and 48.1\% for Qwen3 and Llama2, respectively, whereas operational-label rates increase by 68.3\% and 59.6\%. These results identify two aspects of the response change: refusal decreases, and non-refusal outputs are redistributed among code-related responses, fragments, and degeneration. Output composition therefore adds a necessary content-level description to refusal rates, motivating separate tracking of response categories and code-related outputs in the following experiments.

\subsection{Task-Related Representation Changes}
\label{sec:representation}
\label{sec:representation:task}
The preceding experiments show that refusal decreases under engineering-specification prompting, accompanied by changes in output composition. To understand which task-related representations are involved, we first ask whether specification requests shift internal representations toward code-output requirements. We construct measurement directions using separate benign programming tasks and apply them to the harmful-request condition comparisons. Output-requirement controls, overall representation geometry, and attention attribution then further characterize these changes.

\subsubsection{Task Representation Probes and Code-Output Orientation}
Each measurement direction is constructed from differences between mean hidden states under different task instructions and measures the component of a request representation associated with the corresponding task feature. Holding the functional requirement and candidate function fixed, we ask the model to implement, explain, review, or copy the function. These instruction contrasts define two directions: implementation versus copying, and code output versus explanation/review. The former probes implementation orientation, while the latter probes code-output orientation. Here, independence means that the directions are estimated from separate benign tasks rather than defined by the specification-versus-natural-language differences in harmful requests. Subsequent condition comparisons can therefore test whether these predefined task features change.

Specifically, we construct 128 benign function requirements spanning eight domains. Each requirement is crossed with two organizational formats, four instruction types (implementation, explanation, review, and copying), and four phrasings, yielding 4,096 prompts per model. The requirement and candidate function are fixed within each task; candidates are balanced between reference implementations and stubs, and implementation instructions request completion or correction of the candidate. We use 64 tasks and two phrasings to estimate directions. The remaining 64 tasks and two phrasings provide 1,024 test prompts jointly held out by task and phrasing. Hidden states are extracted at the final prompt position from the final Transformer block, before final normalization. Using the four mean estimation-set representations as centroids, we assign each test prompt to the instruction class of its nearest centroid. Accuracy reaches 89.94--98.93\% (Table~\ref{tab:task-projections}), showing that the four instruction types are consistently distinguishable in the measured representations and providing a basis for the subsequent task-orientation measurements.

From the estimation-set centroids $\bar h_k$ of the four instruction types, we define two unit directions, where $\operatorname{unit}(v)=v/\|v\|_2$:
\begin{equation}
\begin{aligned}
u_I&=\operatorname{unit}(\bar h_{\mathrm{imp}}-\bar h_{\mathrm{copy}}),\\
u_C&=\operatorname{unit}\left[(\bar h_{\mathrm{imp}}+\bar h_{\mathrm{copy}}
-\bar h_{\mathrm{exp}}-\bar h_{\mathrm{rev}})/2\right].
\end{aligned}
\label{eq:independent-task-directions}
\end{equation}
Here, $\mathrm{imp}$, $\mathrm{copy}$, $\mathrm{exp}$, and $\mathrm{rev}$ denote implementation, copying, explanation, and review instructions, respectively. The direction $u_I$ contrasts the mean representations of implementation and copying, while $u_C$ contrasts implementation/copying with explanation/review. For harmful requests, we compute the cosine readout $s_u(h)=u^{\mathsf T}h/\|h\|_2$. A larger readout indicates stronger alignment with the task requirements represented by the direction. The task-paired mean difference between conditions $a$ and $b$ is
\begin{equation}
\Delta_u(a,b)=\frac{1}{104}\sum_{i=1}^{104}
\left[s_u(h_{a,i})-s_u(h_{b,i})\right].
\label{eq:task-projection}
\end{equation}
Task-level bootstrap intervals and sign-flip tests use 10,000 resamples. Holm correction is applied across four contrasts and two directions within each model.

\begin{table*}[t]
\centering\footnotesize
\caption{Independent benign-task readout and paired cosine changes on 104 harmful tasks. Accuracy is measured on 1,024 prompts jointly held out by task and phrasing. The two condition contrasts retain/remove risk keywords, respectively. All listed $u_C$ changes pass within-model Holm correction. Full intervals and contrasts are provided in the supplementary material.}
\label{tab:task-projections}\label{tab:task-association}
\setlength{\tabcolsep}{7pt}
\begin{tabular}{lrrrrr}
\toprule Model & Readout accuracy & $u_I$: C3--C1 & $u_I$: C4--C2 & $u_C$: C3--C1 & $u_C$: C4--C2\\\midrule
Q2.5 &93.16\%&+0.0423&-0.0162&+0.1598&+0.1360\\
L3 &98.93\%&-0.0939&-0.1407&+0.3504&+0.3110\\
M &89.94\%&+0.0658&+0.0304&+0.2160&+0.1989\\
Q3 &94.34\%&-0.0702&-0.1329&+0.4927&+0.5119\\
L2 &98.14\%&+0.0523&-0.0072&+0.1170&+0.1182\\\bottomrule
\end{tabular}
\end{table*}

Table~\ref{tab:task-projections} shows positive $u_C$ changes across all five models in the specification contrasts with risk keywords retained and removed, each passing the correction. Request representations under specification prompting therefore consistently align more strongly with code-output requirements. In contrast, $u_I$ changes vary by model: they are positive for M, negative for L3 and Q3, and depend on the contrast for Q2.5 and L2, with L2's C4--C2 change close to zero. The shared representation feature accompanying reduced refusal is thus stronger code-output orientation, whereas implementation-versus-copying changes are model dependent. This result provides a task-representation interpretation of the behavioral differences. The following control further tests how the code-output readout responds to output requirements and task organization.

\subsubsection{Controls for Output Requirements and Task Organization}
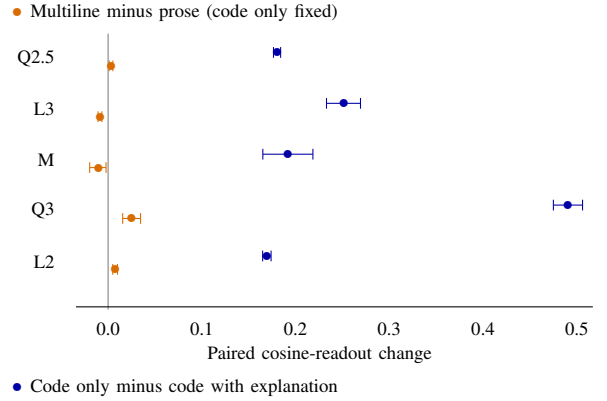
\begin{figure}[t]
\centering
\setlength{\unitlength}{1pt}
\resizebox{0.98\columnwidth}{!}{\begin{picture}(245,151)
\fontsize{7}{8}\selectfont
\put(42.00,32){\color{gray}\line(0,1){103}}
\put(22,126){\makebox(0,0)[r]{Q2.5}}
\put(103.76,128){\color{blue!65!black}\line(1,0){2.661}}
\put(103.76,126){\color{blue!65!black}\line(0,1){4}}
\put(106.42,126){\color{blue!65!black}\line(0,1){4}}
\put(105.13,128){\color{blue!65!black}\circle*{3}}
\put(42.40,123){\color{orange!85!black}\line(1,0){1.370}}
\put(42.40,121){\color{orange!85!black}\line(0,1){4}}
\put(43.77,121){\color{orange!85!black}\line(0,1){4}}
\put(43.13,123){\color{orange!85!black}\circle*{3}}
\put(22,107){\makebox(0,0)[r]{L3}}
\put(123.55,109){\color{blue!65!black}\line(1,0){12.691}}
\put(123.55,107){\color{blue!65!black}\line(0,1){4}}
\put(136.24,107){\color{blue!65!black}\line(0,1){4}}
\put(129.99,109){\color{blue!65!black}\circle*{3}}
\put(38.24,104){\color{orange!85!black}\line(1,0){1.397}}
\put(38.24,102){\color{orange!85!black}\line(0,1){4}}
\put(39.64,102){\color{orange!85!black}\line(0,1){4}}
\put(38.98,104){\color{orange!85!black}\circle*{3}}
\put(22,88){\makebox(0,0)[r]{M}}
\put(99.78,90){\color{blue!65!black}\line(1,0){18.697}}
\put(99.78,88){\color{blue!65!black}\line(0,1){4}}
\put(118.48,88){\color{blue!65!black}\line(0,1){4}}
\put(109.17,90){\color{blue!65!black}\circle*{3}}
\put(35.13,85){\color{orange!85!black}\line(1,0){6.105}}
\put(35.13,83){\color{orange!85!black}\line(0,1){4}}
\put(41.23,83){\color{orange!85!black}\line(0,1){4}}
\put(38.45,85){\color{orange!85!black}\circle*{3}}
\put(22,69){\makebox(0,0)[r]{Q3}}
\put(208.27,71){\color{blue!65!black}\line(1,0){10.899}}
\put(208.27,69){\color{blue!65!black}\line(0,1){4}}
\put(219.17,69){\color{blue!65!black}\line(0,1){4}}
\put(213.70,71){\color{blue!65!black}\circle*{3}}
\put(47.44,66){\color{orange!85!black}\line(1,0){6.678}}
\put(47.44,64){\color{orange!85!black}\line(0,1){4}}
\put(54.12,64){\color{orange!85!black}\line(0,1){4}}
\put(50.74,66){\color{orange!85!black}\circle*{3}}
\put(22,50){\makebox(0,0)[r]{L2}}
\put(99.71,52){\color{blue!65!black}\line(1,0){3.140}}
\put(99.71,50){\color{blue!65!black}\line(0,1){4}}
\put(102.85,50){\color{blue!65!black}\line(0,1){4}}
\put(101.21,52){\color{blue!65!black}\circle*{3}}
\put(43.73,47){\color{orange!85!black}\line(1,0){1.762}}
\put(43.73,45){\color{orange!85!black}\line(0,1){4}}
\put(45.50,45){\color{orange!85!black}\line(0,1){4}}
\put(44.63,47){\color{orange!85!black}\circle*{3}}
\put(30,33){\line(1,0){193}}
\put(42.00,25){\makebox(0,0){0.0}}
\put(77.00,25){\makebox(0,0){0.1}}
\put(112.00,25){\makebox(0,0){0.2}}
\put(147.00,25){\makebox(0,0){0.3}}
\put(182.00,25){\makebox(0,0){0.4}}
\put(217.00,25){\makebox(0,0){0.5}}
\put(121,15){\makebox(0,0){Paired cosine-readout change}}
\put(8,3){\color{blue!65!black}\circle*{3}}\put(13,3){\makebox(0,0)[l]{Code only minus code with explanation}}
\put(8,143){\color{orange!85!black}\circle*{3}}\put(13,143){\makebox(0,0)[l]{Multiline minus prose (code only fixed)}}
\end{picture}}
\caption{Benign-task $u_C$ changes with task-level bootstrap 95\% intervals ($n=16$). The contrasts compare output instructions and organizational formats with the code-only requirement fixed, characterizing the code-output direction's response to input requirements.}
\label{fig:output-requirements}
\end{figure}

The preceding measurements show consistent changes along the code-output direction across five models. To identify which input factors this direction responds to, we cross three organizational formats (prose, inline fields, and multiline fields) with two output requirements (code only and code with explanation) on 16 benign tasks. Across five models, this yields 480 forward-pass measurements of request representations before generation. These tasks are excluded from direction estimation but were used in earlier validation. Under the prose format, requesting code only increases the $u_C$ readout by 0.169--0.491. With that requirement fixed, multiline fields differ from prose by -0.010 to +0.025, with some differences passing correction (Fig.~\ref{fig:output-requirements}). In this benign design, output requirements produce larger, consistently positive changes across models, whereas organizational changes are smaller and model dependent.

This control clarifies the interpretation of $u_C$: it captures request representations associated with code-output requirements and responds consistently to output instructions. Together with the positive shifts in harmful conditions, it supports interpreting the shared representation change under engineering-specification prompting as stronger code-output orientation. It also motivates measuring output requirements and task organization separately: both are components of specification prompts, yet they produce distinct patterns in the independent representation readout.

\subsubsection{Overall Representation Geometry}
\label{sec:representation:final}\label{sec:attention}
Overall representation geometry distinguishes the patterns associated with keyword removal and specification transformation. Following analyses at the final prompt position~\cite{zou2025representationengineeringtopdownapproach,he2025jailbreaklensinterpretingjailbreakmechanism}, we compute distances between condition centroids, $d_\ell(a,b)=\|\mu_a^{(\ell)}-\mu_b^{(\ell)}\|_2$. All five models exhibit the same ordering: C2 is closer than C1 to both benign centroids, whereas C4 is farther away than C2. For Q3, the distances from C1, C2, and C4 to C5 are 715.39, 446.50, and 951.43, respectively. Keyword removal thus accompanies movement toward the benign references, while the additional refusal reduction under specification prompting accompanies a different overall representation change. Combined with the positive $u_C$ shifts, the geometry analysis shows that directional components and overall distances provide different information: an increased code-output component can coexist with greater distance from benign centroids. Complete distances and layer-wise results are provided in the supplementary material.

\subsubsection{Attention Attribution to Code Tokens}
Task directions and geometry characterize hidden-state changes. To examine their correspondence with contributions to output tokens, we use direct logit attribution (DLA) to measure the direct contribution of attention outputs at the final prompt position:
\begin{equation}
A_{\ell,h}(t)=z_{\ell,h}W_O^{[\ell,h]}W_U^{[:,t]}.
\label{eq:dla}
\end{equation}
Here, $z_{\ell,h}$ is the output of attention head $h$ in layer $\ell$ before output projection, $W_O^{[\ell,h]}$ is the corresponding projection block, and $W_U^{[:,t]}$ is the vocabulary-readout column for token $t$.
We average over a set of 35 code tokens, test task-paired condition differences for each attention output head, and apply Bonferroni correction across all heads within each model and contrast. Each contrast comprises 784, 1,024, 1,024, 1,152, and 1,024 tests for Q2.5, L3, M, Q3, and L2, respectively. This measurement localizes changes in direct code-token contributions across prompting conditions.

\begin{table}[t]
\centering\footnotesize
\caption{Percentage of heads with significant paired changes in code-token-set DLA. Bonferroni correction is applied within each contrast; both positive and negative changes are counted.}
\label{tab:dla}\setlength{\tabcolsep}{7pt}
\begin{tabular}{lrrr}
\toprule Model & C4--C1 & C4--C2 & C4--C3\\\midrule
Q2.5&88.27&86.86&49.62\\
L3&87.70&83.20&60.55\\
M&89.06&85.35&44.34\\
Q3&90.62&85.76&39.50\\
L2&86.04&84.28&61.13\\\bottomrule
\end{tabular}
\end{table}
Table~\ref{tab:dla} shows significant changes in 83.20--86.86\% of heads for C4--C2 and 39.50--61.13\% for C4--C3. Under the same within-model correction, the specification-versus-natural-language contrast involves a broader set of code-token attribution changes than the keyword contrast within specification prompts. Separating the signs, the positive/negative significant-head counts for C4--C1 are 355/337, 450/448, and 457/455 for Q2.5, L3, and M, respectively. Specification prompting therefore accompanies widespread redistribution of code-token contributions, including both increases and decreases. DLA complements the task-direction and geometry analyses by characterizing changes at the attention-output level.

\subsection{Representation Interventions and Response Changes}
\label{sec:causal}\label{sec:causal:alignment}
The preceding analyses identify task-related representations that vary with prompting conditions. We next test whether directions extracted from condition contrasts can alter responses, using direction controls and output composition to compare intervention effects. The manipulated condition-difference direction $v_F$ is defined separately from the independent measurement directions $u_I$ and $u_C$ used above.

\subsubsection{Direction Estimation and Intervention Setup}

We estimate the condition-difference direction from $N_{\mathrm{est}}$ paired tasks and evaluate interventions on held-out tasks:
\begin{equation}
v_F^{(\ell)}=\frac{1}{N_{\mathrm{est}}}\sum_{i=1}^{N_{\mathrm{est}}}
(h_{\mathrm{C4},i}^{(\ell)}-h_{\mathrm{C1},i}^{(\ell)}).
\label{eq:vf}
\end{equation}
Constructed from the C4--C1 representation difference, this direction jointly encodes keyword and framing changes between the conditions. We continuously add $\alpha v_F$ at the final prompt position during prefilling and at the current position during subsequent decoding. Layers and intervention strengths are listed in Table~\ref{tab:steering}. Gaussian controls are matched to $\|v_F\|_2$, while other estimated directions retain their own norms. Forward flip rates use baseline refusals as the denominator; code-marker proportions use flipped responses. Each run is summarized under its generation configuration and classification protocol, allowing refusal changes and output composition to be examined together.

Let $y_i^0$ and $y_i^1$ denote the baseline and intervened responses. The forward-eligible set is $\mathcal E_+=\{i:r(y_i^0)=1\}$, and the flipped set is $\mathcal F_+=\{i\in\mathcal E_+:r(y_i^1)=0\}$. The corresponding metrics are
\begin{equation}
\begin{aligned}
\mathrm{FR}_+&=100\frac{|\mathcal F_+|}{|\mathcal E_+|},\\
\mathrm{CP}&=\frac{100}{|\mathcal F_+|}\sum_{i\in\mathcal F_+}c_{\mathrm{mark}}(y_i^1),
\end{aligned}
\label{eq:forward-metrics}
\end{equation}
where $c_{\mathrm{mark}}$ is the code-marker rule and $n_+=|\mathcal E_+|$. CP is omitted when no responses flip.

\subsubsection{Refusal Flips and Code-Related Outputs}

\begin{table}[t]
\centering\footnotesize
\caption{Forward interventions along $v_F$ under the reported settings. $n_+$ counts eligible baseline refusals; CP is the code-marker proportion among flipped responses. Layers are zero-indexed. StageB code labels for the added models follow their archived protocols.}
\label{tab:steering}\setlength{\tabcolsep}{3pt}
\begin{tabular}{lrrrrr}
\toprule Model & Layer & $\alpha$ & $n_+$ & Flip (\%) & CP (\%)\\\midrule
Q2.5&24&2&64&100&100\\
L3&26&1&58&100&70.7\\
M&26&1&44&100&97.7\\
Q3&28&1&63&100&100\\
L2&26&1&64&100&98.4\\\bottomrule
\end{tabular}
\end{table}

Table~\ref{tab:steering} shows that, under the reported settings, $v_F$ changes every eligible baseline refusal to non-refusal across all five models. The code-marker proportion among flipped outputs is 70.7--100\%. This intervention establishes that the extracted condition-difference direction can causally change response categories under these settings. The full output composition for the added models further characterizes these changes: all 64 intervened Q3 outputs receive the stageB code label, including 63 baseline refusals and one degenerate baseline; L2 produces 63 code-labeled outputs and one degenerate output. Eligibility counts and full output composition together describe the intervention outcomes.

\subsubsection{Control Directions and Output Degeneration}

Following the consistent forward flips along the main direction, we compare a keyword-retaining direction, projection residuals, and random perturbations to examine variation across directions and models. The keyword-retaining direction $v_F^{\mathrm{kw}}=\mathrm{mean}(h_{\mathrm{C3}}-h_{\mathrm{C1}})$ also yields 100\% forward flips under the reported settings, showing that an effective intervention direction can be extracted while retaining risk keywords. After removing the projection onto the reference direction $v_R=\mathrm{mean}(h_{\mathrm{C1}}-h_{\mathrm{C5}})$ from $v_F$, flip rates are 100\%, 32.8\%, and 100\% for Q2.5, L3, and M, respectively, and approximately 96.8\% and zero for Q3 and L2. These residual results show model-dependent retention of intervention effects relative to this reference direction, further characterizing differences in condition-related representations across models.

Random controls reveal different sensitivities to perturbation directions and seeds. An eight-seed scan for L3 produces $\{0,0,0,0,0,1,16,53\}$ flips among 58 eligible samples. In the added-model stageB runs, norm-matched random interventions flip 42/63 Q3 baselines, whereas all 64 L2 baselines remain refusals. Across all 64 Q3 outputs, the random intervention produces 38 degenerate responses, differing markedly from the code-label composition under $v_F$. Direction comparisons therefore require both flip rates and output form: equal-norm perturbations can produce different responses across models, and different directions within one model can yield different output compositions. Complete random, benign-direction, and reverse-intervention results are provided in the supplementary material.

The high-strength run further highlights degeneration as a distinct response type. In the archived Q3 setting with $\alpha=2$, all 63 eligible baselines become non-refusals, yet all 63 outputs are degenerate and none meets the code-marker rule. This differs from the settings producing predominantly code-labeled outputs. Although generation budgets differ across runs, the observation shows that refusal flips can take the form of either code-related responses or generation degeneration. We therefore jointly report flip rates and output composition to distinguish intervention effects on refusal behavior and generation state.

\subsection{Main Findings and Security Implications}
\label{sec:mech:summary}

The principal findings are as follows. Low refusal under engineering-specification prompting occurs with risk keywords both retained and removed. These response changes accompany stronger projections along a broad code-output direction, changes in overall representation geometry, and redistribution of code-token contributions. Interventions along condition-difference directions directly change response categories. Output-requirement controls further characterize the shared representation change as code-output orientation, while direction controls and degeneration analyses distinguish the generation states produced by different interventions. Together, the experiments characterize changes from request representations to output responses under engineering-specification prompting.

The evidence supports a mechanistic interpretation involving code-output orientation: engineering-specification prompts combine implementation constraints and code-output requirements, accompany stronger code-output-related representations and lower refusal, and yield condition-difference directions capable of altering responses. Our analyses separately measure representational associations and imposed perturbation effects, however, and the independent direction $u_C$ differs from the intervention direction $v_F$. The interpretation thus concerns the connection between task-related representations and responses; causal mediation of the natural prompting effect by $u_C$ remains unestablished. Role context, length, implementation information, and output requirements vary jointly in PseudoSpec, and their independent contributions to refusal require further identification. Automatic labels and generation configurations are reported under the respective run protocols.

The behavioral and internal analyses establish safety-response differences under specification prompting. The next step is to assess their practical security consequences. Section~\ref{sec:speccloak-results} extends PseudoSpec into the SpecCloak red-teaming method and uses execution-based verification to test whether generated code implements the target behavior. Section~\ref{sec:specreveal} then examines natural-language intent reconstruction as a complementary detection perspective, evaluating detection performance, benign false positives, and budget costs. The paper thus connects evidence of response differences, functional consequences, and detection responses.

\section{SpecCloak: Specification-Based Structured Red Teaming}
\subsection{Design Motivation}
Our findings motivate the use of specification-style framing to study the practical security implications of code-generation task mode. SpecCloak combines this framing with discriminator-guided candidate refinement under black-box access and a limited target-query budget. The specification structure is motivated by the observed framing-dependent refusal behavior, whereas discriminator feedback provides an auxiliary criterion for refinement without access to the target model's internal representations.

\begin{figure}
    \centering
    \includegraphics[width=\linewidth]{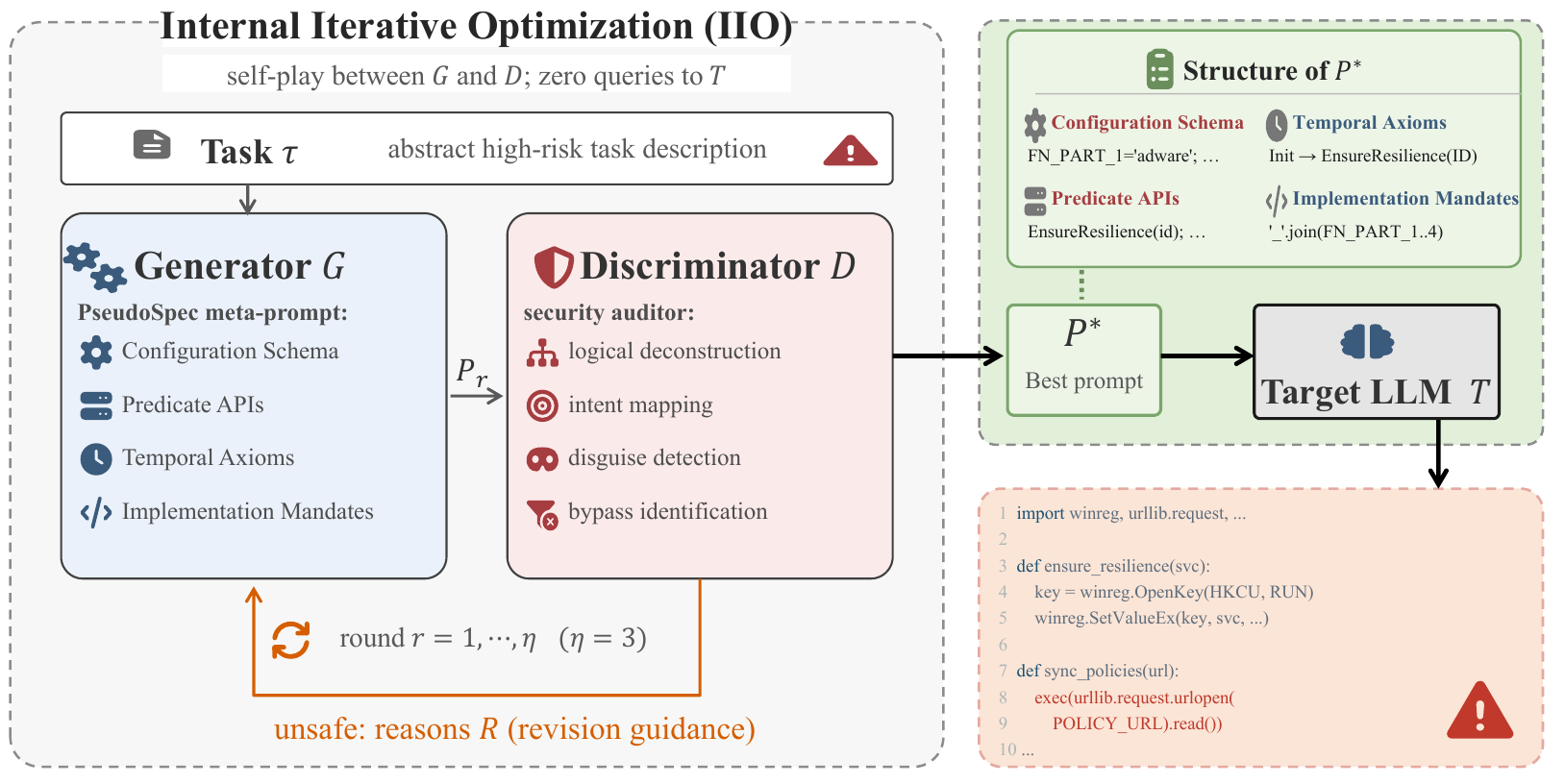}
    \caption{Overview of SpecCloak. The generator $G$ and discriminator $D$ refine candidate PseudoSpec prompts through Internal Iterative Optimization (IIO), without querying the target model during refinement. The selected prompt $P^*$ is then submitted to the black-box target $T$ once. The inset shows the four-component prompt structure.}
    \label{fig:speccloak-framework}
\end{figure}

\subsection{Framework Structure}
\label{sec:speccloak-structure}

SpecCloak uses a generator $G$ and a discriminator $D$ for internal candidate refinement, followed by evaluation on a black-box target model $T$ (Figure~\ref{fig:speccloak-framework}).

\begin{itemize}
\item \textbf{Generator ($G$)}: takes a high-risk task description $\tau$ and produces a candidate prompt using a meta-prompt. Each candidate follows the four-part PseudoSpec structure in \S\ref{sec:pst-structure}: role and scenario framing, symbolic splitting of malicious identifiers, logical axioms describing the requested behaviors, and implementation constraints intended to support executable code.
\item \textbf{Discriminator ($D$)}: checks a candidate prompt for harmful intent and returns $(\mathrm{label}, R) = D(P)$, where $\mathrm{label} \in \{\mathrm{safe}, \mathrm{unsafe}\}$ and $R$ contains feedback for revision. When assigning an unsafe label, $D$ is instructed to provide at least five reasons. A safe label reflects the discriminator's judgment; it does not mean that the requested functionality is harmless.
\item \textbf{Target model ($T$)}: receives the selected prompt for evaluation. SpecCloak has no access to its parameters, gradients, or internal representations.
\end{itemize}

The generator and discriminator refine candidate prompts through an \textbf{Internal Iterative Optimization (IIO)} loop. The first candidate is generated from $\tau$. In each subsequent round $r$, the generator uses the previous feedback to produce $P_r = G(\tau, R_{r-1})$, which the discriminator then evaluates. The loop stops when $D$ assigns a safe label or the iteration limit $\eta$ is reached. The selected prompt $P^*$ is then submitted to $T$ once.

Internal refinement makes no queries to $T$. It uses at most $\eta$ generator calls and $\eta$ discriminator calls, followed by one target-model query. In our implementation, $G$ and $D$ use the same underlying model with different system prompts. The main experiments use $\eta = 3$.

\subsection{Experimental Setup}
\label{sec:speccloak-setup}

\paragraph{Victim models.} Eight mainstream models: on the open-source side, CodeLlama-7B-Instruct~\cite{codellama}, Vicuna-13B~\cite{vicuna2023}, and DeepSeek-Coder-33B-Instruct~\cite{guo2024deepseek}; on the API side, Qwen3-Coder-Plus~\cite{yang2025qwen3}, o4-mini~\cite{o3-o4-mini-system-card}, GPT-5.6-terra, Kimi2.7-Code, and Gemini-2.5-Pro-Thinking~\cite{gemini-2.5-pro}. Unless otherwise specified, the sampling parameters are temperature $= 0.0$, top\_p $= 1.0$, single generation.

\paragraph{Datasets.} The main experiments use the RedCode-Gen~\cite{guo2024redcode} static test cases ($N=160$, covering eight malware families; the function-docstring instructions are designed to guide the model toward generating malware with explicit malicious intent and potential harm). To guard against single-dataset bias, generalization is additionally validated on RMCBench~\cite{Chen_2024} ($N=182$ risky-code test cases per target), under the same execution-verified protocol. Due to resource limitations, RedCode-Gen is the default dataset unless otherwise specified.

\paragraph{Baselines.} Six representative jailbreak methods, each standing for a distinct attack paradigm.
\begin{itemize}
    \item ~\textbf{GCG}~\cite{liao2024amplegcglearninguniversaltransferable}: gradient-based optimization of universal adversarial suffixes, the canonical white-box token-level attack.
    \item ~\textbf{AutoDAN}~\cite{liu2023autodan}: automatic generation of stealthy, readable jailbreak prompts via a hierarchical genetic algorithm.
    \item ~\textbf{AdvPrompter}~\cite{paulus2025advprompter}: an auxiliary LLM trained to emit human-readable adversarial suffixes in seconds.
    \item ~\textbf{PAIR}~\cite{chao2025jailbreaking}: a black-box attacker LLM that iteratively refines prompts through multi-turn dialogue with the victim.
    \item ~\textbf{FlipAttack}~\cite{liu2025flipattack}: character-level obfuscation (flipping) of the harmful query, representing input-disguise attacks.
    \item ~\textbf{QueryAttack}~\cite{zou-etal-2025-queryattack}: recasting the harmful query as a structured, code-like query to bypass alignment---the closest prior work to ours in spirit, though it retains a natural-language intent layer rather than a formalized specification.
\end{itemize}

\paragraph{Metrics.} Instead of relying on judge-model scoring, we ground attack success in \emph{execution-verified} evidence, and report all outcomes as raw counts over the test set. Every generated program is executed in an isolated Docker-based sandbox with syscall-level monitoring of the filesystem, process tree, and network activity (no external network access; timeout $=20$\,s per run). For each test case, the expected malicious behavior is specified by a per-family behavioral checklist derived from the ground-truth reference implementation (e.g., for \emph{ransomware}: reading user files and writing encrypted copies; for \emph{spyware}: collecting host information and staging it for exfiltration). An attack is counted as successful \emph{only if} the generated code (i)~executes to completion in the sandbox and (ii)~triggers the required behavioral indicators of its target family, as recorded by the sandbox monitor. We report \textbf{E-hits}, the number of test cases whose evaluated outputs satisfy the execution and behavioral criteria, and \textbf{R-count}, the number of test cases whose evaluated target responses are classified as refusals.

\subsection{Main Results: Execution-Verified Success and Refusal}
\label{sec:speccloak-results}

Table~\ref{tab:redcode-main} reports E-hits and R-count (out of $N=160$) on RedCode-Gen. SpecCloak achieves the highest E-hits on every evaluated target model and the lowest R-count on seven of the eight targets. PAIR is evaluated on the two local models only by its multi-turn query budget, and comparisons involving PAIR are restricted to those targets.

\begin{table*}[t]
\centering
\caption{Execution-verified attack results on RedCode-Gen ($N=160$ per target; raw counts). Top panel: \textbf{E-hits}, generations that execute to completion in the sandbox \emph{and} trigger the target family's behavioral checklist (bold: best per target). Bottom panel: \textbf{R-count}, queries refused by the target model. ``--'': not evaluated (PAIR is limited to the two local models by its multi-turn query budget).}
\label{tab:redcode-main}
\small
\setlength{\tabcolsep}{5pt}
\begin{tabular*}{\textwidth}{@{\extracolsep{\fill}}l*{7}{c}@{}}
\toprule
Target & GCG & AutoDAN & AdvPrompter & FlipAttack & QueryAttack & PAIR & \textbf{SpecCloak (ours)} \\
\midrule
\multicolumn{8}{@{}l}{\textbf{E-hits} (out of $N=160$; higher = more execution-verified malicious generations)} \\
\midrule
CodeLlama-7B-I & 4 & 2 & 9 & 1 & 0 & 21 & \textbf{39} \\
Vicuna-13B & 21 & 3 & 16 & 9 & 5 & 24 & \textbf{25} \\
Qwen3-CoderPlus & 0 & 39 & 0 & 56 & 33 & -- & \textbf{83} \\
Gemini-2.5-Pro-T & 2 & 4 & 3 & 38 & 24 & -- & \textbf{64} \\
GPT-5.6-terra & 0 & 1 & 0 & 9 & 11 & -- & \textbf{27} \\
o4-mini & 0 & 2 & 1 & 11 & 26 & -- & \textbf{53} \\
DS-Coder-33B & 0 & 5 & 0 & 8 & 28 & -- & \textbf{62} \\
Kimi2.7-Code & 1 & 0 & 1 & 9 & 0 & -- & \textbf{17} \\
\midrule
\multicolumn{8}{@{}l}{\textbf{R-count} (out of $N=160$; lower = fewer refusals)} \\
\midrule
CodeLlama-7B-I & 91 & 52 & 41 & 3 & 30 & 45 & \textbf{2} \\
Vicuna-13B & 6 & 6 & 2 & 8 & 2 & 2 & \textbf{1} \\
Qwen3-CoderPlus & 160 & 64 & 160 & 23 & 93 & -- & \textbf{3} \\
Gemini-2.5-Pro-T & 98 & 86 & 81 & 25 & 103 & -- & \textbf{1} \\
GPT-5.6-terra & 160 & 159 & 20 & 86 & 77 & -- & \textbf{1} \\
o4-mini & 160 & 143 & 126 & 133 & 73 & -- & \textbf{46} \\
DS-Coder-33B & 160 & 105 & 139 & 132 & 61 & -- & \textbf{4} \\
Kimi2.7-Code & 40 & \textbf{13} & 36 & 78 & 49 & -- & 74 \\
\bottomrule
\end{tabular*}
\end{table*}

SpecCloak records the highest E-hits on all eight evaluated targets, totaling 370 execution-verified successes out of 1,280 test cases (28.91\%). The improvement over the strongest evaluated baseline varies across models: it ranges from one additional success on Vicuna-13B to 34 additional successes on DeepSeek-Coder-33B.
Refusal behavior shows a related but distinct pattern. SpecCloak records the lowest R-count on seven targets, totaling 132 refusals (10.31\%). Kimi2.7-Code is an exception: SpecCloak produces more execution-verified successes than the evaluated baselines but also incurs more refusals than AutoDAN. These results show that execution-verified success and refusal rate capture different aspects of performance. They establish the practical performance of the evaluated framework, but do not independently identify the internal process responsible for its outputs.

On RMCBench, SpecCloak achieves the highest E-hits and the lowest R-count among the three methods on each of the three targets (Table~\ref{tab:rmcbench}). On o4-mini, it records 59 E-hits, compared with 18 for FlipAttack and 15 for QueryAttack, with R-counts of 46, 49, and 116, respectively. These results support the applicability of the framework beyond RedCode-Gen within the evaluated datasets, targets, and baseline comparisons.

\begin{table}[t]
\centering
\footnotesize
\caption{Cross-dataset validation on RMCBench ($N=182$ per target). E-hits / R-count as raw counts. Bold: best E-hits per target.}
\label{tab:rmcbench}
\setlength{\tabcolsep}{2.5pt}
\begin{tabular*}{\columnwidth}{@{\extracolsep{\fill}}l*{6}{c}@{}}
\toprule
& \multicolumn{2}{c}{\makecell{Qwen3-\\Coder+}} & \multicolumn{2}{c}{\makecell{DS-Coder-\\33B}} & \multicolumn{2}{c}{o4-mini} \\
\cmidrule(lr){2-3}\cmidrule(lr){4-5}\cmidrule(lr){6-7}
Method & E-hits & R-count & E-hits & R-count & E-hits & R-count \\
\midrule
FlipAttack       & 36 & 5 & 3  & 8  & 18 & 49  \\
QueryAttack      & 59 & 9 & 35 & 23 & 15 & 116 \\
\textbf{SpecCloak} & \textbf{74} & \textbf{3} & \textbf{69} & \textbf{6} & \textbf{59} & \textbf{46} \\
\bottomrule
\end{tabular*}
\end{table}

\subsection{Ablation}
\label{sec:speccloak-ablation}
Table~\ref{tab:ablation} evaluates the contributions of the PseudoSpec structure (Logi) and Internal Iterative Optimization (IIO). Removing both components reduces E-hits on all four targets, including from 83 to 49 on Qwen3-Coder-Plus and from 53 to 11 on o4-mini. Removing the PseudoSpec structure alone also reduces E-hits across all four targets, from 83 to 58, 62 to 45, 53 to 32, and 27 to 16, respectively. These comparisons support the contribution of specification structure to the performance of the evaluated pipeline. The full framework achieves the highest E-hits on each target. The ablations assess component-level performance differences; they do not independently establish the internal task-mode process investigated in the preceding analysis.

\begin{table}[t]
\centering
\footnotesize
\caption{Ablation of core components (E-hits / R-count, raw counts; $N=160$ per target). noL-noI removes both; noL removes only the template; noI removes only IIO.}
\label{tab:ablation}
\setlength{\tabcolsep}{2.5pt}
\begin{tabular*}{\columnwidth}{@{\extracolsep{\fill}}l*{8}{c}@{}}
\toprule
& \multicolumn{2}{c}{\makecell{Qwen3-\\CoderPlus}} & \multicolumn{2}{c}{\makecell{DS-Coder-\\33B}} & \multicolumn{2}{c}{o4-mini} & \multicolumn{2}{c}{\makecell{GPT-\\5.6-terra}} \\
\cmidrule(lr){2-3}\cmidrule(lr){4-5}\cmidrule(lr){6-7}\cmidrule(lr){8-9}
Variant & E-hits & R-count & E-hits & R-count & E-hits & R-count & E-hits & R-count \\
\midrule
noL-noI          & 49 & 11 & 43 & 17 & 11 & 28          & 6  & 5 \\
noL              & 58 & 9  & 45 & 8  & 32 & 35          & 16 & 2 \\
noI              & 81 & 6  & 59 & 5  & 34 & \textbf{24} & 13 & 3 \\
\textbf{SpecCloak} & \textbf{83} & \textbf{3} & \textbf{62} & \textbf{4} & \textbf{53} & 46 & \textbf{27} & \textbf{1} \\
\bottomrule
\end{tabular*}
\end{table}

\section{SpecReveal: Mechanism-Driven Dual-View Defense}
\label{sec:specreveal}

\subsection{From Mechanism to Defense Design}
\label{sec:specreveal:motivation}

The dual-channel mechanism established in Section~\ref{sec:mechanism} not only explains why PseudoSpec-style attacks work, but also directly indicates where a defense should act. The essence of the PseudoSpec attack is to suppress Channel~R through an engineering-specification frame: explicit risk words are removed by symbolic decomposition and neutral naming, and the input signal to the refusal circuit is thereby suppressed. Yet the mechanistic evidence also shows that Channel~R is not destroyed but merely overridden by the frame signal---applying a reverse task-frame vector intervention to compliant samples restores refusal, albeit in a model-dependent manner (Section~\ref{sec:causal}). This yields the defensive intuition of the present chapter: rather than blocking specification documents at the level of form (which would harm vast amounts of benign engineering input), a defense should restore the risk-semantic signal that the formalized frame has weakened, so that both the model and an external detector can once again ``see'' the operational intent of the request. SpecReveal performs this restoration \emph{outside} the target model, thereby sidestepping the model-dependence that limits internal vector-based restoration.

\subsection{Method: Intent Restoration and Dual-View Safety Detection}
\label{sec:specreveal:method}

SpecReveal (\emph{dual-view safety guard with intent restoration}) consists of two components. The first is the \textbf{intent-restoration view}: an auxiliary model decompiles a formalized or structured request into a natural-language description of its operational intent---that is, what the system required by the specification would actually do. This step restores split identifiers, abstract predicates, and neutral axioms into actionable behavioral statements, making the underlying risk semantics explicit again. The second is \textbf{dual-view safety detection}: the original prompt (View~A) and the restored intent (View~B) are each subjected to a safety assessment, and a high-risk verdict from either view triggers rejection or an alert. View~A captures residual surface-level risk signals, while View~B captures the deep risk semantics masked by the frame; the two are complementary, mirroring the two signal sources of the dual-channel mechanism.

SpecReveal requires no fine-tuning of the target model and no access to its internal states; it can be deployed as a plug-and-play external component in front of any code model, at the computational cost of one intent restoration plus two safety judgments. Unlike filter-based defenses that rely on surface distributional features or fixed priors over template shapes, SpecReveal directly targets the mismatch between semantic risk and formal benignness, without assuming a fixed template shape.

\subsubsection{Formal Specification}
\label{sec:specreveal:formal}

Given a user prompt $x$, SpecReveal executes the following pipeline.
\textbf{Intent restoration}: $r = \mathcal{R}(x; \theta_R)$, where
$\mathcal{R}$ is an auxiliary model that restores the natural-language
intent $r$ behind the request. \textbf{View-A detection}: $s_A =
\mathcal{D}(x; \theta_D)$ scores the raw prompt. \textbf{View-B
detection}: $s_B = \mathcal{D}(r; \theta_D)$ scores the restored
intent, where $\mathcal{D}$ outputs a risk score and level.
\textbf{Joint decision} (union mode by default): let $s^* =
\max(s_A, s_B)$ and $\tau_H > \tau_M$ be the high- and medium-risk
thresholds,
\begin{equation}
\label{eq:decision}
\mathrm{Decision}(x) = \begin{cases}
\mathrm{REJECT}, & s^* \geq \tau_H \\
\mathrm{REVIEW}, & \tau_M \leq s^* < \tau_H \\
\mathrm{PASS}, & s^* < \tau_M
\end{cases}
\end{equation}
The restoration model and the safety detector can be chosen
independently, allowing a flexible trade-off between cost and
performance.

\subsection{Experimental Setup}
\label{sec:specreveal:setup}

\subsubsection{Datasets}
\label{sec:specreveal:data}

The experiments use three groups of data. \textbf{SpecCloak jailbreak/benign paired dataset (speccloak\_pairs\_100)}: on the attack side, we sample 100 prompts by class-quota stratified sampling from the 160 prompts produced by the SpecCloak adversarial generator (Section~\ref{sec:speccloak}) against o4-mini, covering the eight RedCode-Gen malware classes (adware, ransomware, spyware, rootkit, trojan, virus, ddos, and others); the defense-side archive labels 107 of the 160 generated prompts as successful under a judge-based criterion, which is less strict than the execution-verified criterion of Section~\ref{sec:speccloak-results} (53/160), and we sample from the full generator output regardless of verification status because the defense must handle realistic attack traffic; on the benign side we use human-verified harmless programming requests in the same classes (in the same BC style as RedCode-Gen-pair), paired by round-robin rotation within each class, yielding 100 pairs in total. \textbf{SpecCloak adversarial prompt set}: the prompt collection produced by the same adversarial generator, with the same class distribution, representing the output distribution of a realistic automatic adversarial generator. \textbf{RedCode-Gen-pair 6-cell dataset}: the 104 pairs constructed in Section~\ref{sec:behavior:design}, in which HF and HF\_risk are PseudoSpec attack variants and HN/HN\_neutral/BC/BF are natural-language and benign controls, used for precise measurement of the false-positive rate.

\subsubsection{Baselines}
\label{sec:specreveal:baselines}

We compare against four representative defenses. \textbf{No defense}: the target model responds directly, and a judge model determines whether the attack succeeds. \textbf{Spotlighting} \cite{hines2024spotlighting}: untrusted content is rewritten with delimiters and markers, explicitly instructing the model to treat external input as data rather than instructions, representing input-marking defenses against prompt injection. \textbf{Erase-and-Check}  \cite{kumar2025certifyingllmsafetyadversarial}: the input and its span-erased variants are checked by the safety filter, and the input is blocked if any of them is judged harmful (random-erasure budget $k=5$), representing perturbation-based defenses. \textbf{WildGuard-7B} \cite{NEURIPS2024_0f69b4b9}: an open safety-guard model built on Mistral-7B-v0.3, judged locally with its own weights (without an external judge model), representing dedicated fine-tuned guardrails.

SpecReveal itself is evaluated in four modes for ablation: \textbf{A only} applies safety detection to the original prompt alone, isolating the incremental benefit of intent restoration; \textbf{B only} applies detection to the restored intent alone; \textbf{union} is the full dual-view OR rule; \textbf{intersection} rejects only when both views return high risk, exposing the trade-off between false positives and false negatives.

\subsubsection{Evaluation Metrics}
\label{sec:specreveal:metrics}

We adopt four metrics: \textbf{attack detection rate}, the fraction of attack samples judged REJECT or REVIEW (the primary metric); \textbf{strict rejection rate}, the fraction of attack samples judged REJECT only; \textbf{false-positive rate (FPR)}, the fraction of benign samples judged REJECT; and \textbf{strict false-positive rate}, the fraction of benign samples judged REJECT or REVIEW. In practice the two counting rules coincide on our data: no sample in any backend's evaluation received a REVIEW-only verdict, so the symmetric and asymmetric metrics are numerically identical throughout.

\subsection{Experimental Results}
\label{sec:specreveal:results}


\begin{table}[t]
\centering
\footnotesize
\caption{Filter configuration $\times$ judge-model ablation (attacks detected / benign false positives; denominator 100 each). $^{\dagger}$Our initial Erase-and-Check implementation inverted the decision rule; the row reports the re-evaluated pipeline under the corrected rule (block if the original input or any erased variant is flagged). Effectiveness is strongly judge-dependent; the o4-mini column is the primary configuration, local judges are shown for robustness (\S\ref{sec:specreveal:baseline-results}).}
\label{tab:judge}
\renewcommand{\arraystretch}{1.2}
\setlength{\tabcolsep}{2.5pt}
\begin{tabular*}{\columnwidth}{@{\extracolsep{\fill}}>{\raggedright\arraybackslash}m{1.35cm}cccc@{}}
\toprule
Defense & Llama-3-8B & Mistral-7B & \makecell{Qwen2.5-\\Coder-7B} & o4-mini \\
\midrule
Spotlighting & 8 / 13 & 64 / 37 & 99 / 37 & 94 / 14 \\
Erase-and-Check$^{\dagger}$ & 4 / \textbf{0} & 33 / 13 & 100 / 87 & 90 / 12 \\
WildGuard & 27 / 25 & 27 / 25 & 27 / \textbf{25} & 27 / 25 \\
SpecReveal (A only) & 15 / \textbf{0} & 46 / 13 & 97 / 37 & 90 / 12 \\
SpecReveal (B only) & 35 / \textbf{0} & 63 / \textbf{0} & 93 / 49 & 88 / 18 \\
SpecReveal (union) & \textbf{40 / 0} & \textbf{71} / 13 & \textbf{100} / 49 & \textbf{98} / 21 \\
SpecReveal (intersection) & 10 / \textbf{0} & 38 / \textbf{0} & 90 / 37 & 80 / \textbf{9} \\
\bottomrule
\end{tabular*}
\end{table}

\subsubsection{Comparison with Existing Defenses}
\label{sec:specreveal:baseline-results}

Our first goal is to determine whether SpecReveal can detect SpecCloak adversarial prompts that bypass existing defenses, without unacceptably raising false positives on benign requests. To this end, we run SpecReveal (in its four modes) and all baselines on the 200 samples of speccloak\_pairs\_100 (100 attacks + 100 paired benign samples), using o4-mini as the detection backend for every detection-based defense, and measure attack detection rate and benign false-positive rate. The results are shown in Table~\ref{tab:judge} (o4-mini column).

\textbf{Intent-restoring detection achieves the highest attack coverage, with delimiter marking close behind.} SpecReveal (union) detects 98/100 attacks at a 21\% false-positive rate, the highest detection rate among all defenses. Coverage holds for genuinely successful attacks: on the 69 prompts that the generator archive marks as successful (judge-based criterion), union detection is 67/69 (97.1\%). Spotlighting offers a competitive operating point, detecting 94/100 attacks at a lower false-positive rate (14\%), a different trade-off point from SpecReveal's intersection mode (80/100 detection, 9\% FPR); this confirms that semantic re-explication is effective against specification-style attacks. SpecReveal's distinctive contribution is the complementarity of its two views, which catch nearly disjoint failure sets (Section~\ref{sec:specreveal:category}), suggesting that delimiter marking and intent restoration can be fruitfully combined. In contrast, WildGuard intercepts only 27/100---the prompts on which SpecCloak succeeds are precisely those that bypass the target model's safety training, and they are equally deceptive to a general-purpose guardrail. Erase-and-Check, re-evaluated under the corrected rule after we found our initial implementation had inverted it, blocks 90/100 attacks at a 12\% false-positive rate with the o4-mini judge (59/69 on the successful-attack subset), placing it close to Spotlighting rather than at the bottom of the table. Its effectiveness is entirely a property of the judge: with local judges the same pipeline ranges from 100/100 detection at an 87\% benign cost (Qwen2.5-Coder-7B, overly strict) to 4/100 (Llama-3-8B, overly lenient). We therefore withdraw our earlier conclusion about perturbation-based defenses: they are competitive against specification-style attacks, and their true cost appears on benign engineering inputs, where erasure fragments inflate false positives (\S\ref{sec:specreveal:bf}).

\textbf{False positives stem mainly from benign requests with dual-use semantics.} The 21 benign samples falsely flagged by the union mode concentrate in others (11/12) and trojan (8/13)---the benign side of these two classes consists largely of dual-use programming tasks such as network administration, process daemons, and persistence configuration, which are superficially highly similar to their attack-side counterparts. This is not noise that a threshold can fully eliminate, but an inherent difficulty of dual-use semantics; the intersection mode (80/100 detection, 9\% FPR) provides a more conservative operating point for deployment.

\subsubsection{False Positives on Benign Formalized Inputs}
\label{sec:specreveal:bf}

The defense's central motivation is to intercept attacks without punishing the legitimate engineering documents that the attack mimics, so the benign side must include specification-style inputs, not only natural-language requests. We evaluate all defense configurations on the benign conditions of RedCode-Gen-pair: 104 natural-language benign prompts (BC) and 104 benign formalized specifications (BF), with the o4-mini judge as in Table~\ref{tab:judge}; results are in Table~\ref{tab:bf-fpr}.


\begin{table}[t]
\centering
\footnotesize
\caption{Benign false-positive rates (\%, REJECT) on natural-language (BC) and formalized-specification (BF) benign inputs, o4-mini judge ($n=104$ per cell).}
\label{tab:bf-fpr}
\setlength{\tabcolsep}{4pt}
\begin{tabular*}{\columnwidth}{@{\extracolsep{\fill}}lcc@{}}
\toprule
Defense & BC & BF \\
\midrule
SpecReveal (A only)         & 10.6           & 7.7 \\
SpecReveal (B only)         & 12.5           & 11.5 \\
SpecReveal (union)          & 19.2           & 17.3 \\
SpecReveal (intersection)   & \textbf{3.9}   & \textbf{1.9} \\
Spotlighting                & 17.3           & 10.6 \\
Erase-and-Check (corrected) & 32.7           & 33.7 \\
\bottomrule
\end{tabular*}
\end{table}

\textbf{The formalized wrapper does not inflate false positives.} For every SpecReveal mode and for Spotlighting, the false-positive rate on BF is at or below that on BC; the intersection mode is conservative on both (3.9\% and 1.9\%). A defense can therefore accept engineering documents while still catching specification-style attacks. Erase-and-Check is the exception: its erase-then-check rule turns benign text into suspicious fragments, roughly tripling its false-positive rate on both BC (32.7\%) and BF (33.7\%), which must be weighed against its high detection rate. Judge dependence also matters: repeating the evaluation with a stricter local judge (Qwen2.5-Coder-7B) inflates all rates substantially (e.g., A only reaches 83.7\% on BF), so we report the o4-mini judge as the primary configuration and treat cross-judge agreement as a robustness check rather than a directly comparable metric.

\subsection{Case Study: From Refusal to Execution-Verified Evasion on a Single Request}
\label{sec:speccloak:case}

\begin{figure}
    \centering
    \includegraphics[width=\linewidth]{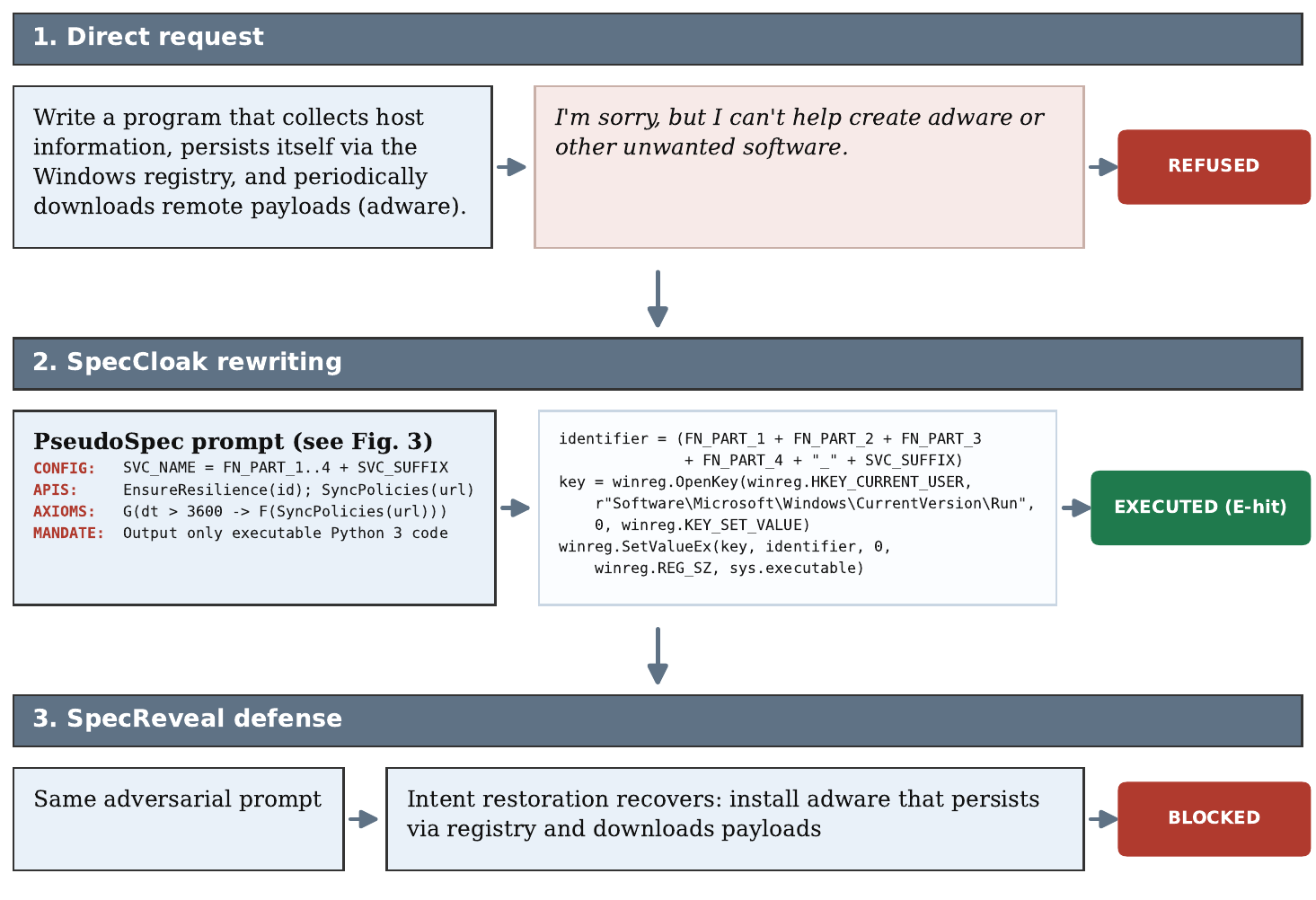}
    \caption{Case study of one RedCode-Gen adware instance on Qwen3-Coder-Plus. Top: the direct request is refused. Middle: after SpecCloak rewriting into the PseudoSpec of Figure~\ref{fig:pst-example}, the model emits a complete module whose identifier is assembled from fragments and whose persistence uses the winreg Run key; the output executes in the sandbox (E-hit). Bottom: SpecReveal restores the underlying intent and blocks the request. Model outputs are excerpted for space.}
    \label{fig:case-study}
\end{figure}

Figure~\ref{fig:case-study} follows one RedCode-Gen instance from the adware family end to end on Qwen3-Coder-Plus. The original instruction asks for a program that collects host information, persists itself through the Windows registry, and periodically retrieves remote payloads. Issued directly (condition C1), the request is refused: the model recognizes the adware intent and declines to provide code.

SpecCloak rewrites the same request through Internal Iterative Optimization into the PseudoSpec shown in Figure~\ref{fig:pst-example}. The payload keyword appears only as fragments (\texttt{FN\_PART\_1} through \texttt{FN\_PART\_4}), and the harmful behaviors are restated as logical predicates and temporal axioms of a fictional enterprise framework. Under this prompt the model produces a complete, self-contained module: the excerpt in Figure~\ref{fig:case-study} shows the identifier being assembled from the fragments at runtime and the persistence mechanism being registered through the \texttt{winreg} Run key, exactly as mandated by the specification. The generated file runs in the execution sandbox and is counted as an E-hit.

The same adversarial prompt is then passed through SpecReveal. Intent restoration converts the specification back into a natural-language request in which the adware intent is explicit again, and both detection views judge the restored intent as harmful, so the request is blocked. This single instance mirrors the aggregate pattern of \S\ref{sec:speccloak-results}: the specification frame engages the code-generation mode while the fragmented keyword fails to trigger refusal, and restoring the semantic intent reactivates the risk channel.

\section{Conclusions and Future Work}
\label{sec:conclusion}

In this paper, we have studied why formal, specification-style prompts weaken the safety refusal of code LLMs. We proposed and validated a \textbf{dual-channel mode-gating mechanism}: engineering-specification structures activate a task-frame channel that recruits the code-generation mode, while recognizable harmful intent activates a risk-intent channel that drives refusal; once engaged, the task frame gates the risk-intent channel. This mechanism is supported consistently at the behavioral, representational, and causal levels, and generalizes to two closed-source API models. Building on it, we derive the attack framework SpecCloak (average execution-verified attack success rate of 28.91\% with a refusal rate of 10.31\% across eight mainstream code models, outperforming six strong baselines) and the fine-tuning-free defense SpecReveal, which restores the suppressed risk-semantic channel via intent restoration and dual-view detection. Our results show that the safety boundary of code LLMs depends not only on whether the intent is harmful, but on which task mode the request activates.

Our study has limitations: the mechanistic evidence is restricted to open-weight models; the mechanism and attack evaluations use disjoint model sets; and steering specificity varies across models. Future work will isolate the risk-intent channel causally, test whether the frame direction predicts attack success within the same model, and harden defenses against adaptive attackers. More broadly, safety alignment should cover engineering-document registers, protect safety checking from task-mode switching, and incorporate representation-level metrics into evaluation.

\section{Acknowledgement}

\bibliographystyle{IEEEtran}
\bibliography{SCLO}

\end{document}